# Mechanical, thermal and optical properties of perovskite borides $R$Rh$_3$B ($R$ = Y, Zr, and Nb)

**M.A. Hossain[1], M.S. Ali[2], A.K.M.A. Islam[2*]**

[1]*Department of Physics, Mawlana Bhashani Science and Technology University, Santosh, Tangail-1902, Bangladesh*

[2]*Department of Physics, Rajshahi University, Rajshahi, Bangladesh*

## ABSTRACT

We report here *ab initio* density functional theory (DFT) calculations of structural, elastic, Peierls stress, thermodynamic and optical properties of $R$Rh$_3$B ($R$ = Y, Zr and Nb) using the plane wave pseudopotential method. The materials possess better ductile behavior in comparison with a selection of layered MAX phases but the anisotropy is strong, particularly in NbRh$_3$B. The Peierls stress, approximately 3-4 times larger than in MAX phases, show that dislocation movement may follow but with much reduced occurrences compared to MAX phases. The temperature and pressure dependence of bulk modulus, specific heats, thermal expansion coefficient, and Debye temperature are calculated for the first time for two of the three compounds using the quasi-harmonic Debye model with phononic effects for elevated temperature and pressure. The obtained results are discussed in comparison to the behavior of other related compounds. Further the features of optical functions obtained for the first time are discussed. The study reveals that the reflectivity is high in the IR-UV regions up to ~ 17.5 eV (YRh$_3$B, ZrRh$_3$B) and 20 eV (NbRh$_3$B), thus showing promise as good coating materials.

*Keywords*: $R$Rh$_3$B; Quasi-harmonic Debye model; Thermodynamic properties; Optical properties

## 1. Introduction

The rare earth rhodium borides $R$Rh$_3$B ($R$ = Y, Zr, and Nb) possess cubic perovskite structure with space group $Pm\bar{3}m$ and form a part of general formula $R$M$_3$X, where $R$ and M rare earth metals and 4$d$ metals, respectively and X is B, C or N. The phases have a great importance due to their interesting properties including high stability and hardness, which make them promising candidates for high-temperature environments, hard coating applications, and cutting tools [1- 6]. Further there was a motivation to search for superconductivity in these compounds which was prompted by the very interesting and quite non-typical superconductivity of MgCNi$_3$ perovskite [7], with $T_c$ ~ 8 K. Among other intermetallic perovskites, similar to MgCNi$_3$, only YRh$_3$B with $T_c$ = 0.76 K [8] was previously reported to be a superconductor: However, since discovery of superconductivity in MgCNi$_3$, no other superconductor has been found so far.

Schaak *et al.* [2] synthesized $R$Rh$_3$B compounds by arc melting of powder. Music *et al.* [3,4] have suggested that $R$M$_3$X compounds may possess an unusual combination of metallic and ceramic properties due to interleaving of high and low electron density layers. The suggestion stems from the similarity of the compounds in the elastic characteristics and electronic structure with the MAX phases, where an early transition metal carbide or nitride is interleaved primarily with an element belonging to IIIA or IVA group [9-11]. These perovskite phases exhibit a combination of different properties, e.g. high stiffness [12-14], plastic deformability [13], good thermal and electrical conductivity [15,16] and resistance to oxidation [17].

As mentioned earlier YRh$_3$B is a superconductor with low $T_c$ [8] and hardness of 8 GPa [18]. The electronic band structure and density of states of superconducting YRh$_3$B have been investigated by Ravindran *et al.* [19] and they observed that the B *p* states and Rh *d* states are completely degenerate from the bottom of the conduction band to Fermi level $E_F$ and this implies that there is a strong covalent bonding existing between B and Rh. Music *et al.* [3] have studied the correlation between electronic structure and mechanical properties of YRh$_3$B. Music and Schneider [20] have also investigated the effect of valence electron concentration on elastic properties of $R$Rh$_3$B. After we completed our present work we have come across a very recent investigation by Litimein *et al.* [21] who have presented the structural, elastic and thermodynamic properties of ScRh$_3$B, LaRh$_3$B and YRh$_3$B.

It is thus evident from the above discussion that there has been a certain amount of experimental and theoretical work performed on rare earth rhodium boride RRh$_3$B compound [8, 20-22]. But there is a dearth of information on other perovskite borides. To the best of our knowledge, thermal and optical properties of $R$Rh$_3$B ($R$ = Zr, Nb) compounds have not yet been discussed in literature. The thermodynamic properties of solids include a variety of properties and phenomena such as bulk modulus, specific heat, thermal expansion coefficient, Debye temperature and so on. The microscopic thermodynamic properties are closely related to the microscopic dynamics of atoms. The specific heat of a material is one of the most important thermodynamic properties indicating its heat retention or loss ability. On the other hand the optical properties of solids provide an important tool for studying energy band structure, impurity levels, excitons, localized defects, lattice vibrations, and certain magnetic excitations. The optical conductivity or the dielectric function indicates a response of a system of electrons to an applied field. Thus there is a need to deal with all these issues which will be covered in the present paper and a discussion and analysis will be made in comparison with results of MAX phases and other relevant compounds.

## 2. Computational methods

The zero-temperature energy calculations have been performed using CASTEP code [23] which utilizes the plane-wave pseudopotential based on density functional theory (DFT). The electronic exchange-correlation energy is treated under the generalized gradient approximation (GGA) in the scheme of Perdew-Burke-Ernzerhof (PBE) [24]. The interactions between ion and electron are represented by ultrasoft Vanderbilt-type pseudopotentials for Y, Zr, Nb, Rh, and B atoms [25]. The calculations use a planewave cutoff energy 400 eV for all cases. For the sampling of the Brillouin zone, 7×7×7 *k*-point grids generated according to the Monkhorst-Pack scheme [26] are utilized. These parameters are found to be sufficient to lead to convergence of total energy and geometrical configuration. Geometry optimization is achieved using convergence thresholds of 5×10$^{-6}$ eV/atom for the total energy, 0.01 eV/Å for the maximum force, 0.02 GPa for the maximum stress and $5\times10^{-4}$ Å for maximum displacement. Integrations in the reciprocal space were performed by using the tetrahedron method with a $k$-mesh of 20 $k$-points in the irreducible wedge of Brillouin zone (BZ). The total energy is converged to within 0.1mRy/unit cell during the self consistency cycle.

To investigate the thermodynamic properties, we employed the quasi-harmonic Debye model, the detailed description of which can be found elsewhere [27]. Through this model, one could calculate the thermodynamic parameters including the bulk modulus, thermal expansion coefficient, specific heats, and Debye temperature etc. at any temperatures and pressures using the DFT calculated energy-volume data at $T = 0$ K, $P = 0$ GPa and the Birch-Murnaghan EOS [28].

## 3. Results and discussion

### 3.1. Mechanical properties

The rare earth transition metal boride perovskites with space group $Pm\bar{3}m$ belong to the cubic systems, the unit cell of which is displayed in Fig. 1. The equilibrium crystal structure of $R$Rh$_3$B is first obtained by minimizing the total energy. The relevant optimized lattice parameters are shown in Table 1 along with measured values obtained by Schaak *et al.* [2]. The calculated values of lattice parameters deviate by 1.4%–5% from the measured values, the largest deviation is found for NbRh$_3$B, which has been reported to be substoichiometric [2].

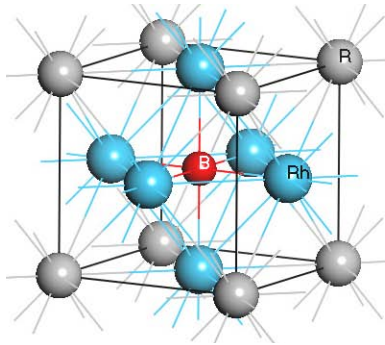

**Fig. 1.** Equilibrium crystal structure of $R$Rh$_3$B.

The elastic constant tensor of perovskite $R$Rh$_3$B phases is reported in Table 1 along with available computed elastic constants. Using the second order elastic constants, the bulk modulus $B$, shear modulus $G$, Young's modulus $E$, and Poisson's ratio $v$ at zero pressure are calculated and illustrated in the table. The dependence of the elastic constants is a very important characterization of the crystals with varying temperature and pressure, which will be dealt with in a later section. We also show Burgers vector $b$ and interlayer distance $d$ between the glide planes as explained in Eq. (1) later.

It is seen from Table 1 that $R$Rh$_3$B phases are characterized by relatively high elastic constants, as evident from the obtained values of Young's modulus and bulk modulus. The ductility of any solid material can be roughly estimated by the ability of performing shear deformation, such as the value of shear-modulus-to-bulk-modulus ratios. According to Pugh's criteria [30], a material should behave in a ductile manner if $G/B < 0$; otherwise, the material is brittle. The calculated $G/B$ ratios as a function of Z($R$) of the perovskites along with a single value from ref. [21] are shown in Fig. 2. As is evident, the $G/B$ ratios (0.46 - 0.21) decrease as Z is increased, and are less than those of selected 312 MAX phases and binary carbide TiC (see [31]). Thus the perovskites under consideration show better ductile behavior in comparison with a selection of layered MAX phases which are mentioned here along with their $G/B$ values within the parentheses [32-34]: $Mo_2GaC$ (0.39), $Nb_2SC$ (0.40, 0.54), $Nb_2InC$ (0.44), $Nb_2AsC$ (0.50), $Nb_2SnC$ (0.59, 0.54), $Ti_2InC$ (0.66, 0.79) $Ti_2InN$ (0.65).

**Table 1.** Calculated lattice parameters $a$ (Å), elastic constants $C_{ij}$ (GPa), bulk moduli $B$ (GPa), shear moduli $G$ (GPa), Young's moduli $E$ (GPa), Poisson's ratio $v$, $b$ (Å) and $d$ (Å) for $R$Rh$_3$B phases.

| Structure | $a$ | $C_{11}$ | $C_{12}$ | $C_{44}$ | $B$ | $G$ | $E$ | $v$ | $b$ | $d$ |
|---|---|---|---|---|---|---|---|---|---|---|
| YRh$_3$B | 4.266 | 333 | 112 | 72 | 186 | 86 | 223 | 0.30 | 4.266 | 2.08 |
| | 4.209[a] | 322[a] | 120[a] | 65[a] | 184[a] | 78[a] | 204[a] | 0.68[a*] | | |
| | 4.22[b] | 339[b] | 108[b] | 67[b] | 183[b] | - | - | - | | |
| | 4.216[c] | - | - | 68[c] | 181[c] | - | - | - | | |
| | 4.207[d] | - | - | - | 208[d] | 96[d] | 250[d] | 0.30[d] | 4.21[h] | 2.10[h] |
| | 4.191[e] | - | - | - | 205[e] | - | - | - | | |
| | 4.22[f] | - | - | - | 177[f] | - | - | - | | |
| | 4.168[g] | - | - | - | - | - | - | - | | |
| ZrRh$_3$B | 4.153 | 400 | 132 | 79 | 222 | 98 | 255 | 0.306 | 4.153 | 2.08 |
| | 4.159[c] | - | - | 69[c] | 225[c] | - | - | - | | |
| | 4.094[g] | - | - | - | - | - | - | - | | |
| NbRh$_3$B | 4.115 | 381 | 161 | 25 | 234 | 48 | 134 | 0.405 | 4.115 | 2.06 |
| | 4.112[c] | - | - | 37[c] | 231[c] | - | - | - | | |
| | 3.922[g] | - | - | - | - | - | - | - | | |

a: [21] * The value of $v$ is erroneously quoted as 0.66 instead of 0.33.
b: [22] Young's and shear moduli are erroneously quoted.
c :[20]
d: [3]
e. [19]
f : [29]
g: [2]

We find that bulk modulus $B$ increases by 28% from186 to 234 GPa, while $C_{44}$ decreases by 65% from 72 to 25 GPa (Table 1). As shown in Fig. 2 the $B/C_{44}$ ratio range from 2.6 to 9.4 is unusually large for ceramics which exceeds MAX phases. Earlier Music and Schneider [20] reported similar behavior of the variation of $B/C_{44}$ values which are also displayed in Fig. 2 for comparison. They explained this behavior using electronic band structure and speculated that this large ratio may indicate unusual plasticity. Here the predominantly covalent-ionic Rh-B layers are interleaved with predominantly metallic Rh-$R$ layers. Thus as observed earlier the large $B$-to-$C_{44}$ ratio is a consequence of weak coupling between the Rh-B and Rh-$R$ layers (giving rise to low $C_{44}$) as well as strong coupling within Rh-B layers (giving rise to a large modulus $B$).

The knowledge of elastic anisotropy is important in diverse applications such as phase transformations, dislocation dynamics, development of microcracks, and other geophysical applications. Recently it has been emphasized that the elastic anisotropy influences the nanoscale precursor textures in shape memory alloys [35] The quantification of the elastic anisotropy is due to Zener [36], which is defined by $A = 2C_{44}/(C_{11}−C_{12})$. According to our calculations the value of $A$ decreases from 0.65 to 0.23 as Z increases (Fig. 2), indicating that NbRh$_3$B has a profound anisotropy compared to other two compounds.

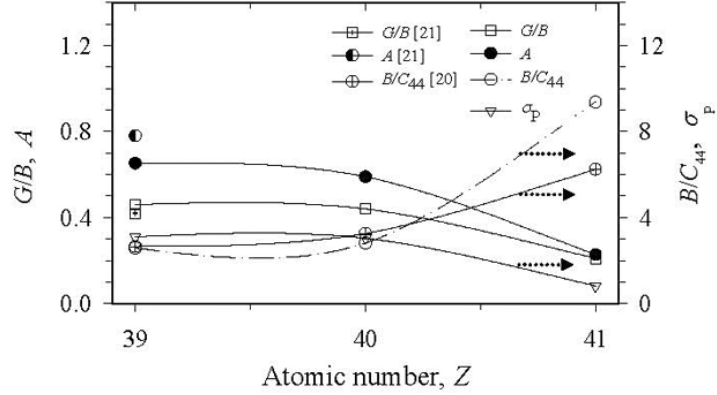

**Fig. 2.** The ratio $G/B$, shear anisotropy $A$, $B/C_{44}$, and $\sigma_P$ of perovskites as a function of atomic number ($Z$) of their rare earth metals $R$.

The initial stress value required to initiate the movement of a dislocation in a glide plane can be estimated using Peierls stress ($\sigma_P$) [37], which is expressed through shear modulus and Poisson ratio as follows:

$$\sigma = \frac{2G}{1-\nu}\exp\left(-\frac{2\pi d}{b\left(1-\nu\right)}\right) \tag{1}$$

where $b$ is the Burgers vector and $d$ is the interlayer distance between the glide planes. For perovskites the Burgers vector $b$ is for $<001>$ plane, and interlayer distance $d$ for $1/2(001)$ plane [38] (see Table 1). The estimated Peierls stress data for perovskites under consideration are shown in Fig. 2. These may be compared with the values of a selection of MAX phases (0.98, 0.86, 0.78, 0.76, 0.71 of Ti$_2$AlC, Cr$_2$AlC, Ta$_2$AlC, V$_2$AlC, Nb$_2$AlC, respectively) [31] and rocksalt binary carbide TiC (19.49). We see that $\sigma_P$(MAX) $< \sigma_P$(perovskites) $< \sigma_P$(binary carbide). Thus it is clear that in the MAX phases dislocations can move, while this is not the case for the binary carbides. The perovskites studied here exhibit an intermediate values of $\sigma_P$, approximately 3-4 times larger than in MAX phases, so that dislocation movement may also occur.

### 3.2. Thermodynamic properties at elevated temperature and pressure

We investigated the thermodynamic properties of $R$Rh$_3$B by using the quasi-harmonic Debye model, the detailed description of which can be found in literature [27]. Here we computed the normalized volume, bulk modulus, specific heats, Debye temperature and volume thermal expansion coefficient at different temperatures and pressures for the first time. For this we utilized $E$-$V$ data obtained from the third-order Birch-Murnaghan equation of state [28] using zero temperature and zero pressure equilibrium values, $E_0$, $V_0$, $B_0$, based on DFT method within the generalized gradient approximation. The thermodynamic properties at finite-temperature and finite-pressure can then be obtained using the model. The non-equilibrium Gibbs function G*($V$; $P$, $T$) can be written in the form [27]:

$$G^*(V; P, T) = E(V) + PV + A_{vib}[\Theta(V); T]$$

(2)

where $E(V)$ is the total energy per unit cell, $PV$ corresponds to the constant hydrostatic pressure condition, $\Theta(V)$ is the Debye temperature, and $A_{vib}$ is the vibrational term, which can be written using the Debye model of the phonon density of states as [27]:

$$A_{vib}(\Theta, T) = nkT\left[\frac{9\Theta}{8T} + 3\ln(1 - \exp(-\Theta/T)) - D\left(\frac{\Theta}{T}\right)\right]$$

(3)

where $n$ is the number of atoms per formula unit, $D(\Theta/T)$ represents the Debye integral.

The function $G^*(V; P, T)$ can now be minimized with respect to volume $V$ to obtain the thermal equation of state $V(P, T)$ and the chemical potential $G(P, T)$ of the corresponding phase. Several other macroscopic properties can also be derived as a function of $P$ and $T$ from standard thermodynamic relations [27]. Here we computed the bulk modulus, Debye temperature specific heats, and volume thermal expansion coefficient at different temperatures and pressures.

Fig. 3(a) shows the temperature dependence of isothermal bulk modulus of $R$Rh$_3$B phases at $P = 0$ GPa. The zero pressure theoretical results for YRh$_3$B due to Litimein *et al.* [21] shown as (+) in the figure are seen to decrease at a faster rate. The bulk modulus increases as $Z$ increases, and as the temperature increases the bulk modulus decreases slowly and the rate of decrease is nearly same for all the compounds under consideration. Further the bulk modulus, signifying the average strength of the coupling between the neighboring atoms, increases with pressure at a given temperature (see inset of Fig. 3 (a)) and decreases with temperature at a given pressure, which is consistent with the trend of volume of the perovskites.

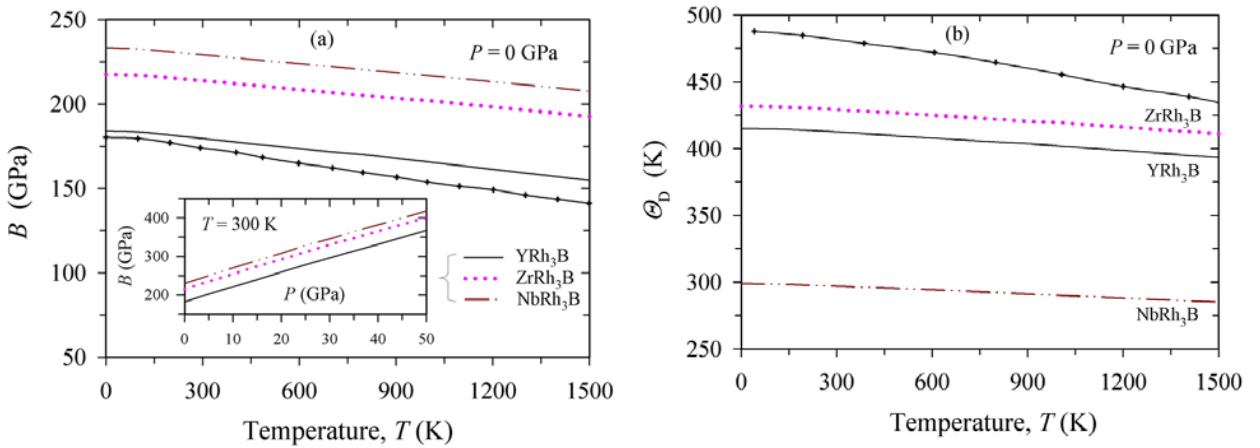

**Fig. 3.** The temperature dependence of (a) bulk modulus and (b) Debye temperature of $R$Rh$_3$B at $P = 0$ GPa in comparison with results (+) for YRh$_3$B [21]. The inset to the left figure shows the pressure dependence of $B$ at 300 K.

Fig. 3 (b) displays temperature dependence of Debye temperature $\Theta_D$ for the three perovskites at zero pressure along with the theoretical result due to Litimein *et al.* [21] for only YRh$_3$B compound. Our $\Theta_D$ value at $P = 0$ GPa, $T = 0$ K for YRh$_3$B is less by 12.5%. This may be due to different calculational methodology used. One observes that $\Theta_D$, smallest for NbRh$_3$B phase, decrease slightly in a non-linear way with temperature for all the perovskites. Further pressure variation of $\Theta_D$ (not shown) shows a non-linear increase. The variation of $\Theta_D$ with pressure and temperature reveals that the thermal vibration frequency of atoms in the perovskites changes with pressure and temperature.

The temperature dependence of constant-volume and constant-pressure specific heat capacities $C_V$, $C_P$ of $R$Rh$_3$B ($R$ = Y, Zr and Nb) are shown in Fig. 4 (a, b). Theoretical results of $C_V$ for YRh$_3$B [21] and MgCNi$_3$ [39] are also plotted for comparison, in addition to room temperature experimental value for the latter compound. The heat capacities increase with increasing temperature, because phonon thermal softening occurs when the temperature increases. The difference between $C_P$ and $C_V$ for $R$Rh$_3$B is given by $C_P - C_V = \alpha_V^2(T) \, BTV$, which is due to the thermal expansion caused by anharmonicity effects. In the low temperature limit, the specific heat exhibits the Debye $T^3$ power-law behavior and at high temperature ($T > 400$ K) the anharmonic effect on heat capacity is suppressed, and $C_V$ approaches the classical asymptotic limit of $C_V = 3nNk_B = 124.7$ J/mol.K for $R$Rh$_3$B ($R$ = Y, Zr and Nb). These results show the fact that the interactions between ions in $R$Rh$_3$B have great effect on heat capacities especially at low temperatures. There is no theoretical or experimental value of specific heat capacities of $R$Rh$_3$B except for YRh3B. Instead we have shown the theoretical and experimental data of superconducting antiperovskite MgCNi$_3$ for comparison with the present calculated result.

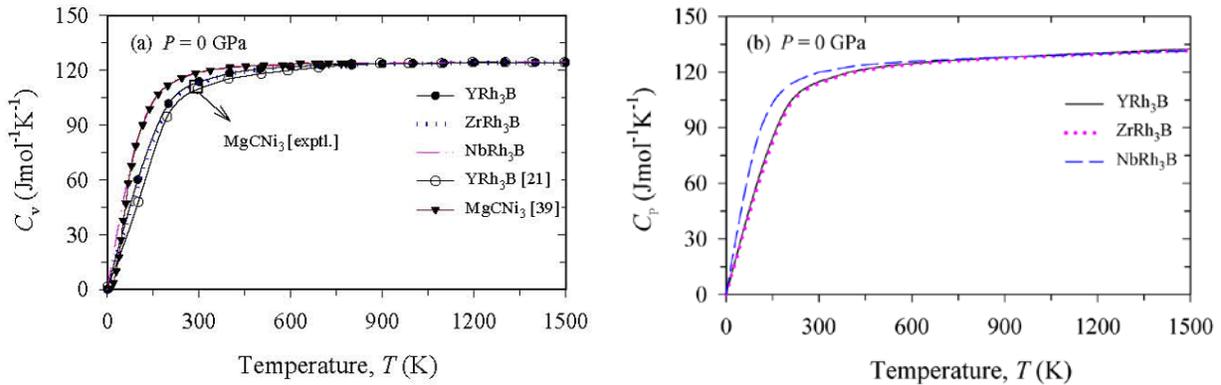

**Fig. 4.** The temperature dependence of (a) constant-volume and (b) constant-pressure specific heat capacities $C_V$, $C_P$ of $R$Rh$_3$B ($R$ = Y, Zr and Nb) in comparison with available theoretical [21, 39] and experimental data [40].

Figs. 5 (a) and (b) show the volume thermal expansion coefficient $\alpha_V$ of $R$Rh$_3$B perovskites as a function of temperature and pressure, respectively. The theoretical results of only YRh$_3$B due to Litimein *et al.* [21] using the full-potential APW + lo method with the mixed basis as implemented in the WIEN2K code are also shown for comparison. There are substantial differences in the values predicted by the two different computational methodologies, particularly above room temperature. The thermal expansion coefficient increases rapidly especially at temperature below 300 K, whereas it gradually tends to a slow increase at higher temperatures. On the other hand at a constant temperature, the expansion coefficient decreases strongly with pressure. It is well-known that the thermal expansion coefficient is inversely related to the bulk modulus of a material.

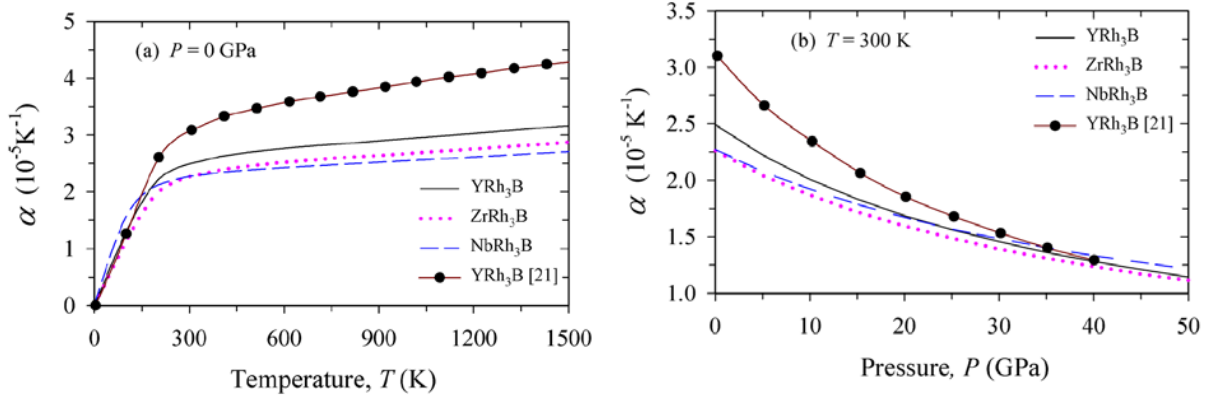

**Fig. 5.** The volume thermal expansion coefficient $\alpha_V$ as a function of (a) temperature and (b) pressure for YRh$_3$B, ZrRh$_3$B and NbRh$_3$B compared with available results.

### 3.3. Optical Properties

Compound having cubic symmetry requires only one dielectric tensor component to characterize the linear optical properties. The imaginary part of the dielectric function $\varepsilon_2(\omega)$ is calculated using the expression given in Ref. [41]. The real part $\varepsilon_1(\omega)$ is derived from the imaginary part by the Kramers-Kronig transform. From the real and imaginary parts of the dielectric function one can calculate other spectra, such as refractive index, absorption spectrum, loss-function, reflectivity and conductivity (real part) using the expressions given in Ref. [41].

Fig. 6 shows the optical functions of $R$Rh$_3$B calculated for photon energies up to 25 eV for polarization vectors [100]. We have used a 0.5 eV Gaussain smearing for all calculations. This smears the Fermi level in such a way that $k$-points will be more effective on the Fermi surface. The calculations only include interband excitations. In metal and metal-like systems there are intraband contributions from the conduction electrons mainly in the low-energy infrared part of the spectra. Electronic structures study and chemical bonding analysis show that considered perovskites are conductors and their bonding characterized by a mixture of ionic-covalent type [19, 20]. It is thus necessary to include this via an empirical Drude term to the dielectric function [42, 43]. A Drude term with plasma frequency 3 eV and damping (relaxation energy) 0.05 eV was used.

The imaginary and real parts of the dielectric function are displayed of Fig. 6 (a, b). It is observed that the real part $\varepsilon_1$ goes through zero from below and the imaginary part $\varepsilon_2$ approaches zero from above at about 18.3, 18.5, and 20.2 eV for YRh$_3$B, ZrRh$_3$B and NbRh$_3$B, respectively. Metallic reflectance characteristics are exhibited in the range of $\varepsilon_1 < 0$. The peak of the imaginary part of the dielectric function is related to the electron excitation. For the imaginary part of $\varepsilon_2$, the peak for < 1.5 eV is due to the intraband transitions. It is clear from the figure that $\varepsilon_2(\omega)$ shows single peak at ~ 2 eV (YRh$_3$B), two peaks at ~ 1.7 and 2.7 eV (ZrRh$_3$B), two peaks at ~ 1.2 and 2.4 eV (NbRh$_3$B). The refractive index and extinction coefficient are illustrated in Fig. 6 (c) and (d). The calculated static refractive index $n(0)$ is found to be 0.6, 0.25, and 3, for YRh$_3$B, ZrRh$_3$B, and NbRh$_3$B, respectively.

Fig. 6 (e) shows the absorption coefficient spectra of the perovskites under consideration. Since the materials have no band gap as evident from band structures, the photoconductivity starts with zero photon energy for each of the phases as shown in Fig. 6 (f). These spectra have several maxima and minima within the energy range studied. The photoconductivities and hence electrical conductivities of materials increase as a result of absorbing photons. The maximum optical conductivity [in (mΩcm)$^{-1}$] of 43.9 (YRh$_3$B), 66.9 (ZrRh$_3$B), and 71.1 (NbRh$_3$B) was observed at photon energy of ~ 2.2, 3.0, and 2.6 eV, respectively.

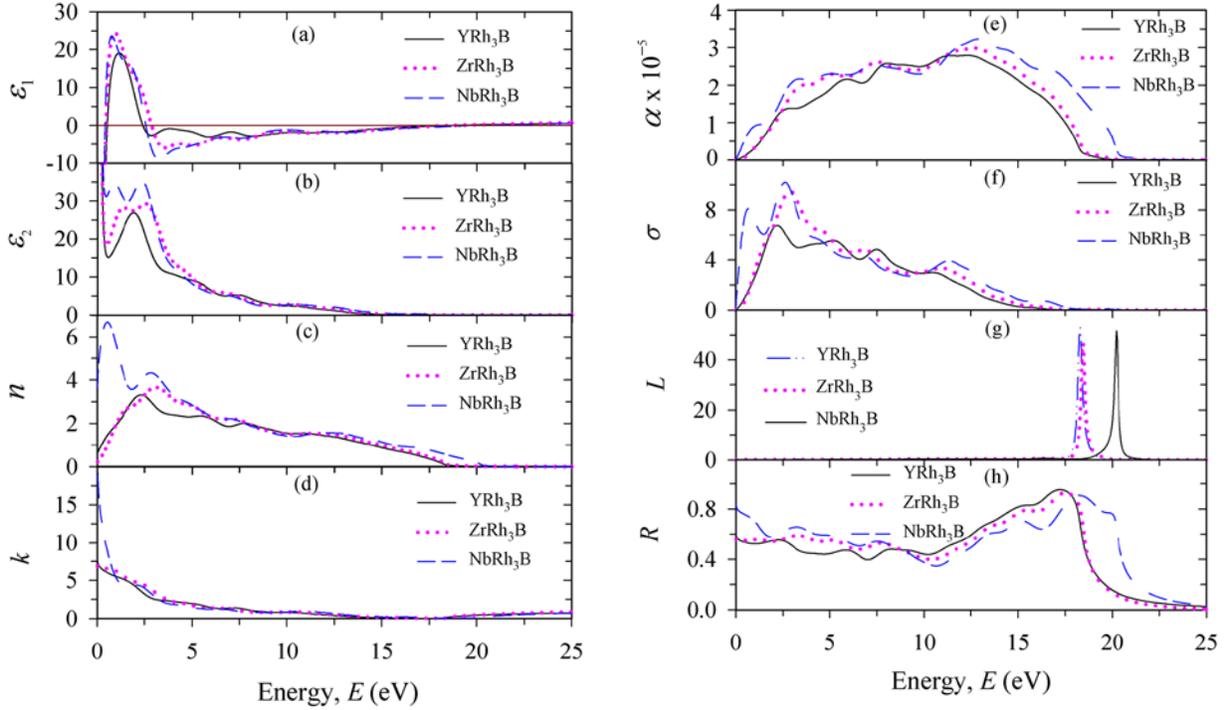

**Fig. 6.** Energy dependent (a) real part of dielectric function, (b) imaginary part of dielectric function, (c) refractive index, (d) extinction coefficient, (e) absorption, (f) real part of conductivity, (g) loss function, and (h) reflectivity of $R$Rh$_3$B along [100] direction.

Fig. 6 (g) and 6 (h) show the electron energy loss spectra $L(\omega)$ and reflectivity spectra $R(\omega)$ as a function of photon energy, respectively of the three perovskites. The function, $L(\omega)$, unfolds the energy loss of a fast electron passing through a material. We see that $L(\omega)$ shows three sharp peaks one for each phase at 18.3, 18.5, and 20.2 eV for YRh$_3$B, ZrRh$_3$B, and NbRh$_3$B, respectively. This peak represents the feature that is associated with plasma resonance, and the corresponding frequency is called bulk plasma frequency. Further these peaks correspond to irregular edges in the reflectivity spectrum (Fig. 6 (h)), and hence an abrupt reduction occurs at these peaks values in the reflectivity spectrum and it correlates with the zero crossing of $\varepsilon_1(0)$ with small $\varepsilon_2(0)$, shown in Fig. 6 (a, b). The reflectivity values of the three perovskites are high between 0 and ~20 eV photon energy, reaching maximum of ~93%. The reflectivity is thus much higher than those in the superconducting antiperovskite MgCNi$_3$ [44]. It implies that $R$Rh$_3$B can all be used as good coating materials in the ultraviolet region.

## 4. Conclusion

First-principles calculations based on DFT have been used to study the mechanical, thermodynamic and optical properties of YRh$_3$B, ZrRh$_3$B and NbRh$_3$B. The materials are elastically anisotropic and show better ductility in comparison with a selection of layered MAX phases but the anisotropy is strong, particularly in NbRh$_3$B. The estimated values of Peierls stress indicate that dislocation movement may follow in the perovskites but with much reduced occurrences compared to MAX phases. The finite-temperature ($\leq 1\,500$ K) and finite-pressure ($\leq 50$ GPa) thermodynamic properties, *e.g.* bulk modulus, specific heats, thermal expansion coefficient, and Debye temperature are all obtained through the quasi-harmonic Debye model, which considers the vibrational contribution, and the results are analysed. The variation of $\Theta_D$ with temperature and pressure reveals the changeable vibration frequency of the particles in $R$Rh$_3$B. The heat capacities increase with increasing temperature, which shows that phonon thermal softening occurs when the temperature increases. From the analysis of optical functions for the polarization vectors [100], it is found that the reflectivity high in IR-UV regions and it implies that $R$Rh$_3$B (Y, Zr and Nb) materials can be used as coating materials.